# Nanoripple formation on GaAs (001) surface by reverse epitaxy during ion beam sputtering at elevated temperature

Debasree Chowdhury, Debabrata Ghose[*]

*Saha Institute of Nuclear Physics, Sector – I, Block – AF, Bidhan Nagar, Kolkata 700064, India*

## Abstract

Self-organized pattern formation by the process of reverse epitaxial growth has been investigated on GaAs (001) surfaces during 1 keV Ar$^+$ bombardment at target temperature of 450$^0$C for a wide range of incident angles. Highly ordered ripple formation driven by diffusion instability is evidenced at near normal incidence angles. Concurrent sample rotation shows that the ripple morphology and its orientation do not depend on the incident beam direction; rather they are determined by the symmetry of the crystal face.

---

*Correspondence: debabrata1.ghose@gmail.com

## 1. Introduction

Molecular beam epitaxy (MBE) has recently been applied to the growth and self assembly of three-dimensional (3D) nanostructures on various semiconductor substrates for a number of applications in the field of electronics and photonics [1, 2]. The origin of this phenomenon is attributed to an intrinsic instability of the surface in the presence of an extra energy barrier, the so-called Ehrlich-Schwoebel (ES) barrier [3, 4] felt by diffusing adatoms at monoatomic steps where they bounce back instead of reaching of lower terraces. As the interlayer mass transport is inhibited, there is a net uphill current perpendicular to the step direction resulting to an unstable surface. Such atomistic mechanism, termed as Villain instability [5], leads to the growth of fascinating patterned morphologies following the symmetry of the crystalline surface.

Another method for nanostructure formation on semiconductor surfaces is the low energy ion beam sputtering (IBS) which is well-known for its cost effectiveness and short production time [6, 7]. Contrary to MBE, surface diffusion barrier and in particular ES barrier does not found to be responsible for the growth of surface instability in IBS of semiconductors. Instead, the surface curvature dependent sputtering [8] plays the dominant role for the development of 3D nanostructures. The main reason behind this is the amorphization of the surface during ion bombardment, which destroys the surface structure and the effects of ES barrier. It is, however, possible to re-activate the ES biasing if one dynamically anneals the disordered surface by keeping the target temperature above its recrystallization temperature. Recent experiments [9 - 12] showed that at elevated target temperatures, i.e. under conditions of rapid damage annealing, ion beam erosion of crystalline semiconductors can show an inverse homoepitaxial growth analogous to MBE. The creation of excess vacancies by sputtering and the maintenance of terrace step structure at high temperature can trigger the action of diffusion bias on surface vacancies so as to destabilize the vicinal surface with respect to the structure formation. In MBE, the active diffusion species is the adatoms which are incorporated at a higher rate at ascending steps leading to the growth of mound-like structures. On the contrary, the IBS produces surface vacancies which are the main diffusing species and have a higher attachment rate at descending steps thereby leading to development of "troughs", i.e. a reverse growth of the surface structure. Such asymmetric kinetics for the attachment of vacancies is found to have stronger destabilization effects compared to the curvature-dependent sputtering under certain bombardment conditions, e.g. at normal ion beam sputtering [9].

In this paper, we report highly ordered ripple formation on GaAs (001) surface under 1 keV $Ar^+$ ion bombardment at sample temperature of $450^0$ C by reverse epitaxy. We have shown experimentally that the pattern morphology at near normal incidence angles is exclusively controlled by surface diffusion dynamics especially the step-edge barriers analogous to MBE. On the other hand, at grazing incidence angles the curvature-dependent sputtering mechanism is found to play the dominant role for the pattern formation.

2. **Experimental**

The experiments were carried out in a high vacuum sputtering chamber with a residual pressure of $10^{-8}$ mbar. The ion beam system provides a broad ion beam of 4 cm diameter from an inductively coupled RF discharge ion source equipped with three graphite grid ion optical system (M/s Roth & Rau Microsystems GmbH, Germany). The substrate holder has the facility of tilting the sample from $0^0 - 90^0$ with respect to the normal to the surface and also has the provision for azimuthal rotation around the surface normal up to the speed of 30 rpm. The GaAs (001) wafers with a miscut of $\pm\, 0.5^0$ were exposed to high purity Ar[+] beam of 1 keV energy and were placed about 33 cm away from the ion source. Care has been taken to prevent surface contamination with metallic impurities from the chamber wall and/or the ion source [13]. The beam current density was about 1000 µA cm$^{-2}$. The sample could be heated up to $500^0$ C by a radiation heater mounted at the chamber top. The temperature was measured by a thermocouple. After irradiation, the sputtered surface was examined *ex-situ* in a Veeco NanoScope IV atomic force microscope (AFM) operating in the tapping mode with Si cantilevers of 10 nm nominal tip radius.

## 3. Results and discussion

Figure 1 shows the AFM images of 1 keV Ar[+] irradiated GaAs (001) surfaces at target temperature $T_g$ of $450^0$C and at incidence angles θ between $0^0$ and $85^0$. The bombarding fluence was fixed at $1 \times 10^{19}$ ions cm$^{-2}$. Development of surface patterns at different angular regime is clearly visible. For comparison, the surface morphology at normal incidence bombardment and at room temperature ($T_g = 20^0 C$) is also presented, which shows a flat stable surface. The rms roughness of bare GaAs was measured to be 0.162 nm and after irradiation at room temperature, it reduces to 0.109 nm, i.e. there is smoothing of the surface. Interesting changes of the surface morphology, however, occurs when the target is irradiated at elevated temperature. Highly ordered periodic ripples are found to develop at normal incidence ($\theta = 0^0$) of ion bombardment [cf., Fig. 1(b)]. The morphology consists of alternate piled-up elongated terraces with nanogrooves in-between running parallel to $\langle 1\bar{1}0 \rangle$ direction. The width of the terraces is quite narrow (a few nm) being along the $\langle 110 \rangle$ direction. With the increase of incidence angle the terrace width increases, but the shape becomes increasingly irregular. Although a large number of topological defects are produced, a clear anisotropy is still observed in the pattern.

For examining the symmetry and the order of the pattern, the fast Fourier transform (FFT) of AFM images are calculated and displayed in Fig. 2. The origin of the reciprocal space locates at the center of each spectrum. The FFT at room temperature irradiation exhibits a broad range of spatial frequencies showing a random structure. For the FFTs at high temperature irradiation and incident angles up to $65^0$, one can see two bright spots located along the horizontal axis with the origin in-between. These spots come from the rippled surface and it displays a two-fold symmetric structure. The spot radius, which is inversely proportional to the correlation length, decreases as the incident angle progresses towards the normal. Also the two spot centers are separating further, indicating the reduction of ripple wavelength. The second order diffraction spots can be seen for bombardment at smaller incident angles, revealing a high degree of order for the ripple pattern at near normal incidence angles.

As the incident angles are increases, the symmetry becomes less and less prominent as the two spots tend to merge and finally, at $70^0$ an isotropic structure is obtained. In the grazing incidence angles, i.e. in the angular region $75^0$ - $85^0$ the topographical pattern is not clearly discernible in real space, but the Fourier space showing an elliptic structure indicates the formation of an anisotropic morphology with the wave vector perpendicular to the incident beam direction independently from the azimuthal orientation of the sample surface.

In order to ascertain that the ripple structure at low incident angles originates from the crystallinity of the surface, we rotate the sample around the surface normal azimuthally at a constant speed of 5 rpm after tilting the surface at various θ angles. This is equivalent to the system of a fixed sample irradiated uniformly by the ion beam from all azimuthal angles. The results are shown in Fig. 3. For each tilted angle, including $0^0$ angle of incidence, we found that the pattern orientation does not change under continuous rotation of the target, i.e. ripple forms with its crest remained align with the <1$\bar{1}$0> crystallographic direction similar to that obtained for the non-rotating target. This clearly indicates that the formation of ripples is determined by the bombarded surface crystallography and not by the ion beam direction with respect to the surface normal. On the other the hand above the angle of incidence $65^0$, the anisotropy is smeared out by rotation and the pattern exhibits an isotropic symmetry as revealed from the corresponding FFT images. This shows that, for the latter case, the pattern formation is mainly dictated by the incident beam direction [13].

To analyze the statistical properties of the developed morphology, we extract the height data from the AFM images (Fig. 1) and evaluate the height-height correlation function defined as $G(r) = \langle (h_i - h_j)^2 \rangle$, where $h_i$ and $h_j$ are heights of the surface at two locations, $i$ and $j$, separated by a distance $r$, and the bracket $\langle \ \rangle$ denotes averaging over pairs of points [14, 15]. For the patterned structures calculations of $G(r)$ along the horizontal direction, shows a number of oscillations after the linear part of the curve. The value of $r$ corresponding to the first minimum is designated as $d$, the dominant in-plane length scale, which quantifies the separation or repeat distance between dominant features of the surface morphology. The surface roughness amplitude $w$ is defined as the value of $\sqrt{G(r)}$ at the first local maximum. Plots of $G(r)$ as a function of $r$ for different ion incidence angles θ are displayed in Fig. 4. Periodic oscillations of $G(r)$ are clearly discernible for bombardment up to the angle of $55^0$ and the wavelength of the ripples evaluated from the position of the first local minimum is found to vary from 40 – 60 nm [Fig. 5(a)]. The roughness amplitude $w$ versus θ [Fig. 5(b)] shows a bell-shaped type variation with a broad maximum in the angular region θ = 45 – $55^0$.

Analysis of local slope distributions provides information about the formation of predominant facets of the ripple structure. The two-dimensional slope distributions extracted from AFM images (Fig. 1) are shown in Fig. 6, evidencing the selection of facet slopes. The projection of $\nabla h$ along $\langle 110 \rangle$ direction peaks at well-defined values that depend on the incident angle θ. The evolution of the selected slopes as a function of θ is plotted in Fig. 7. Globally the slope angle is found to be constant up to the incidence angle of θ ≈ $45^0$, which is attributed to the formation of well-defined facets of the ripple structure. A saturation of the slope around $20^0$ identifies the development of the facet planes corresponding to {114} ($19.47^0$) planes tilted with respect to the (001) surface. In the continuum description of kinetic instabilities, such a slope selection is linked to ES barriers and is determined by the nonlinear current $\vec{j}_{NE}$ that vanishes for a given value of the local slope [16]. For θ > $45^0$, the slope angle decreases showing diminishing strength of the diffusion instability.

The rippling of the surface observed in the angular range $0^0$ - $65^0$, can be understood by a diffusion-biased roughening mechanism in the presence of ES barrier [4, 5]. The supersaturated vacancies generated by sputtering on a high symmetry surface prefer to stick to the lower terraces rather than ascending because of the ES barrier. These result in a net downhill current leading to a surface instability in the form of depressions which subsequently coarsen and lead to the development of 3D morphology. This mechanism has been proposed by Ou et al. [9, 12] to explain the evolution of various types of symmetric pattern on sputtered Si (001), Ge (001), Ge (111), InAs (001) and GaAs (001) surfaces. Irrespective of the bombarded beam direction with respect to the azimuth, the pattern prefers to remain aligned with the direction of the crystallographic axes characterizing the bombarded face

The strength of biased diffusion, however, depends on the vacancy density and the degree of biasing in attachment at descending steps. The transition between diffusive and erosive regime can be induced by changing the ion beam conditions, e.g. the incident angle to suppress or enhance different mechanisms [17]. The ES instability is, found to be weakened above a certain critical angle ∼ $65^0$. At this stage, the instability due to the curvature dependent sputtering is greater and the pattern formation is dominated by the erosion kinetics. For the latter, the ion beam direction dictates the growth of surface structure independently from the crystal orientation. The present results substantiate that obtained from single crystal metal sputtering experiments performed earlier by the group of Valbusa [17, 18].

## 4. Conclusions

We observe instabilities and pattern formation on ion eroded GaAs (001) surface at elevated temperature arising from surface diffusion dynamics analogous to those observed in MBE. Highly ordered defect-free ripples develop at near-normal incidence angles ($\theta \approx 0^0 - 35^0$) by 1 keV $Ar^+$ bombardment at sample temperature of $450^0$ C. As the incident angle is increased up to $65^0$, the ordering of the ripple structure, however, degrades but maintaining its orientation along the <$1\bar{1}0$> direction. Above $65^0$, the pattern vanishes and smoothening of the surface is observed, although in the reciprocal space a clear change of orientation of the surface structure by $90^0$ can be recognized. Moreover, the sample rotation does not alter the overall sputtered-morphology and the pattern orientation so long as the incident angle remains less than $70^0$. Above $70^0$, the rotation induces an isotropic structure. The present results, thus, show that at low angles of bombardment the kinetic instability involves with a diffusion bias vacancy transport, whereas at grazing incidence angles the mechanism of curvature dependent sputtering governs the structure formation.

**Acknowledgements**

One of the authors (D. G.), the Emeritus Scientist, CSIR, thanks CSIR (Grant No. 21(0988)/13/EMR-II dated 30-04-2015), New Delhi for providing financial support.

**Figure Captions:**

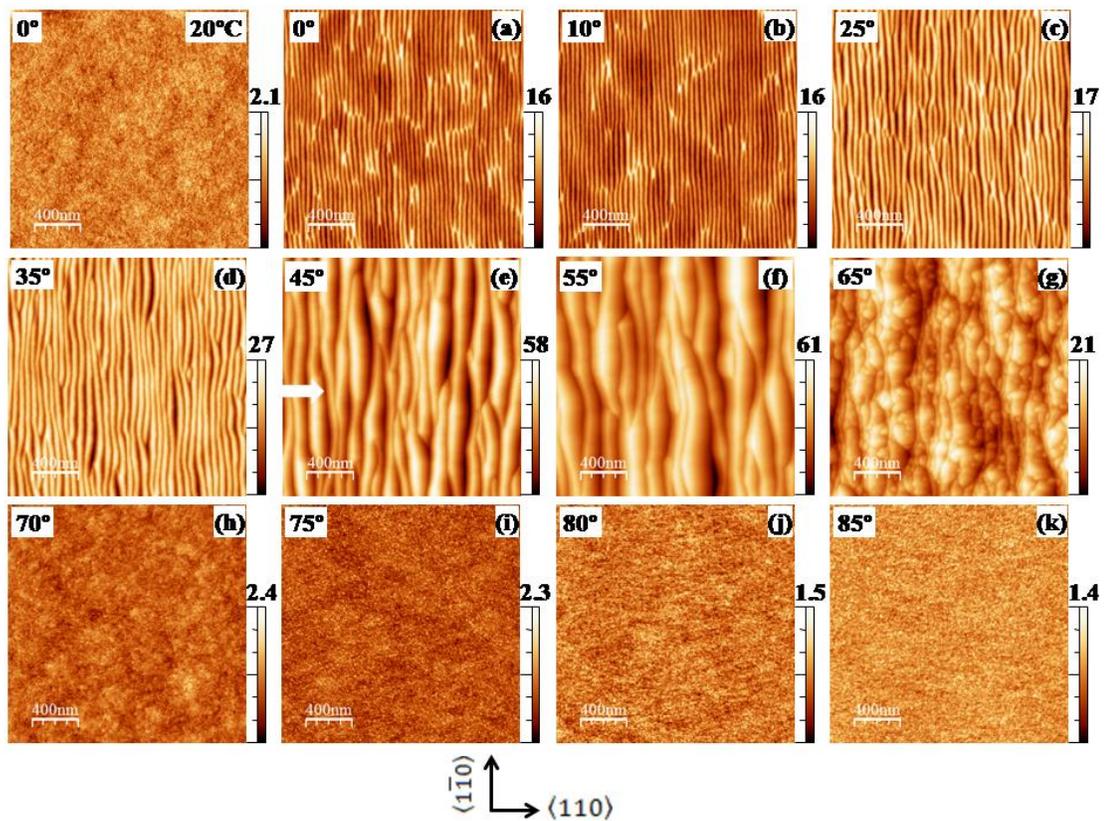

Fig. 1: The AFM topography of 1000 eV Ar$^+$ irradiated GaAs (001) surfaces for a fixed fluence of $1 \times 10^{19}$ ions/cm$^2$ at different ion incidence angles. The sample image (a) is for $T_g$ = 20$^0$ C;

(b) – (k) for $T_g = 450^0$ C. The $\langle 110 \rangle$ and $\langle 1\bar{1}0 \rangle$ crystal directions are marked. The white arrow indicates the projection of the ion beam direction. The vertical length scale is in nm.

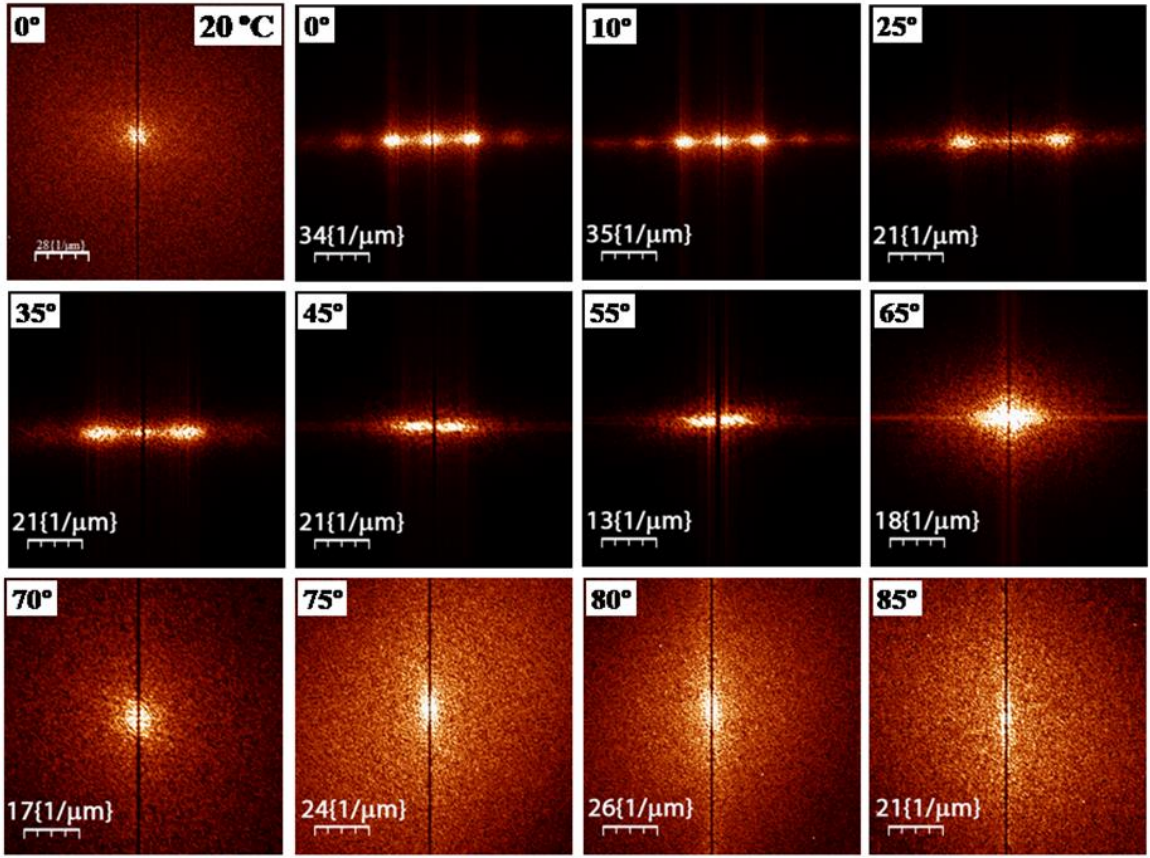

Fig. 2: The First Fourier Transform (FFT) of the AFM images displayed in Fig. 1. For comparison at the top left corner the FFT image at room temperature irradiation is shown.

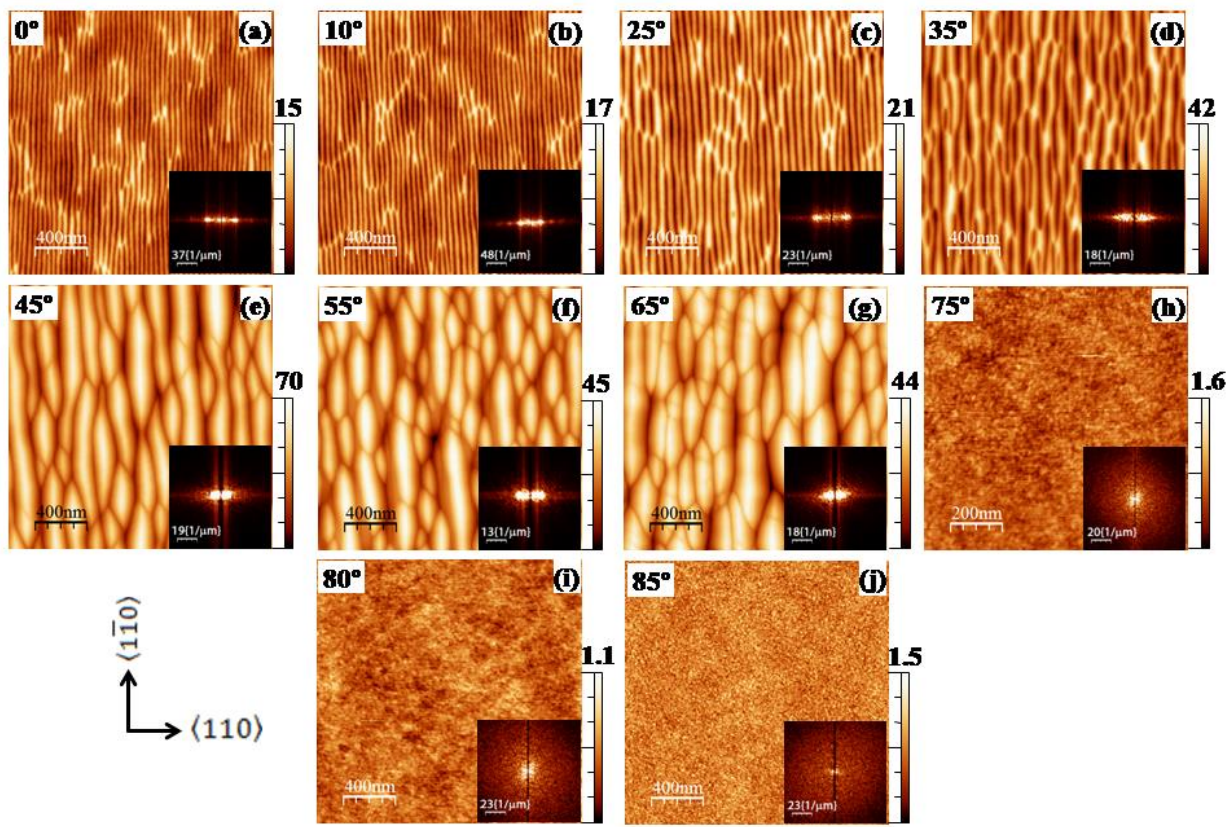

Fig. 3: Result of substrate rotation on the evolution of nanopatterns on GaAs (001) surface at different ion incidence angles for $T_g$ = 450$^0$ C. Energy of Ar$^+$ = 1 keV; substrate rotation = 5 rpm; fluence = 1 × 10$^{19}$ ions/cm$^2$. The vertical length scale is in nm.

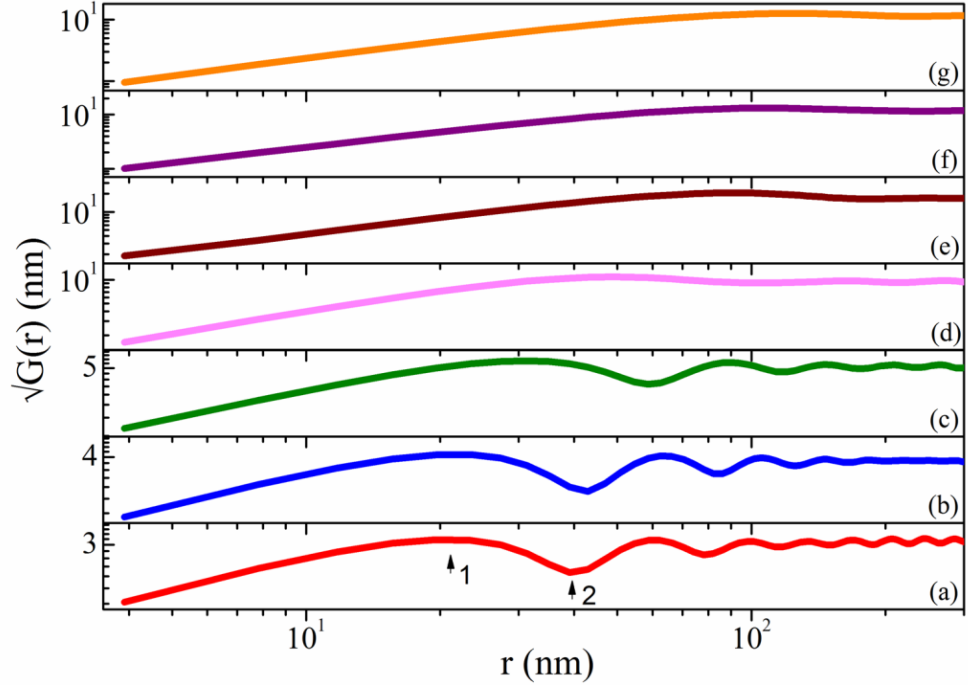

Fig. 4: Calculated height-height correlation functions $\sqrt{G(r)}$ as a function of $r$ for increasing incidence angles: (a) $0°$, (b) $10°$, (c) $25°$, (d) $35°$, (e) $45°$, (f) $55°$, (g) $65°$; energy of $Ar^+$ = 1 keV; $T_g$ of GaAs (001) = $450°$ C; fluence = $1 \times 10^{19}$ ions/cm$^2$. The roughness amplitude and in-plane length scale are indicated by arrows 1 and 2, respectively, for the curve (a).

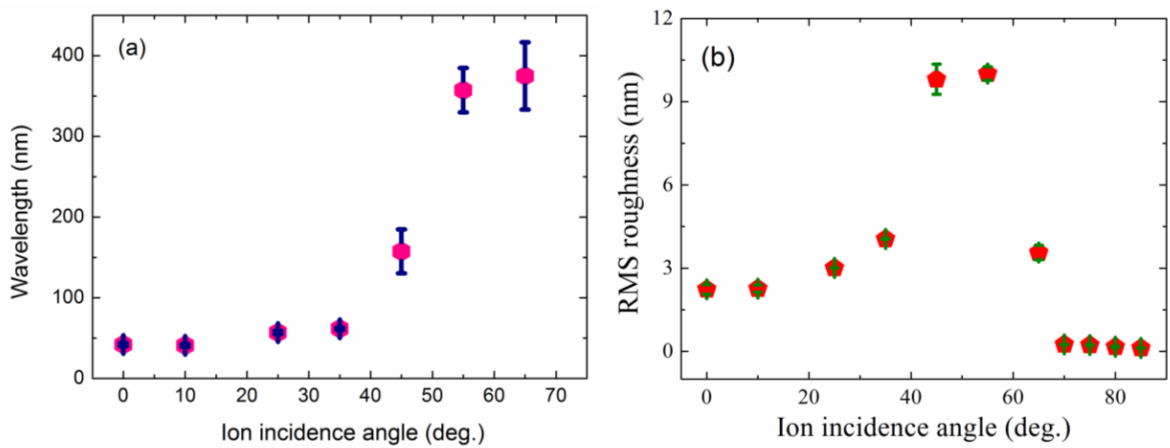

Fig. 5: Evolution of (a) the *rms* roughness and (b) the wavelength from Fig. 4 as a function of ion fluence.

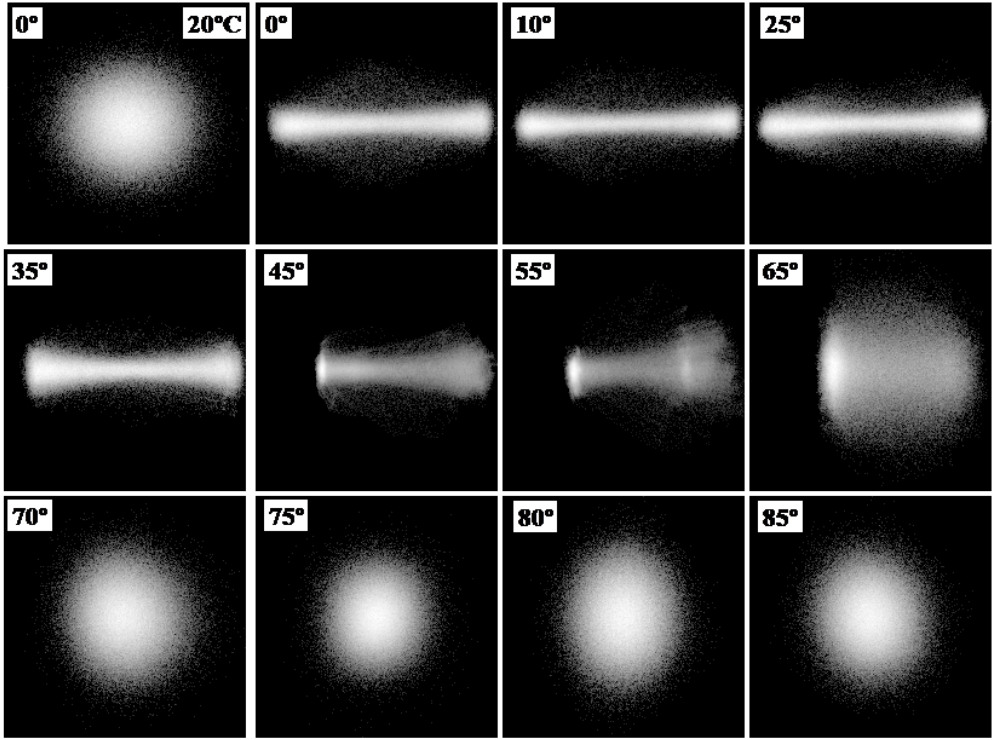

Fig. 6: Slope angle distributions of the AFM images displayed in Fig. 1. For comparison, the slope distribution of the topography at room temperature irradiation is shown at the top left corner.

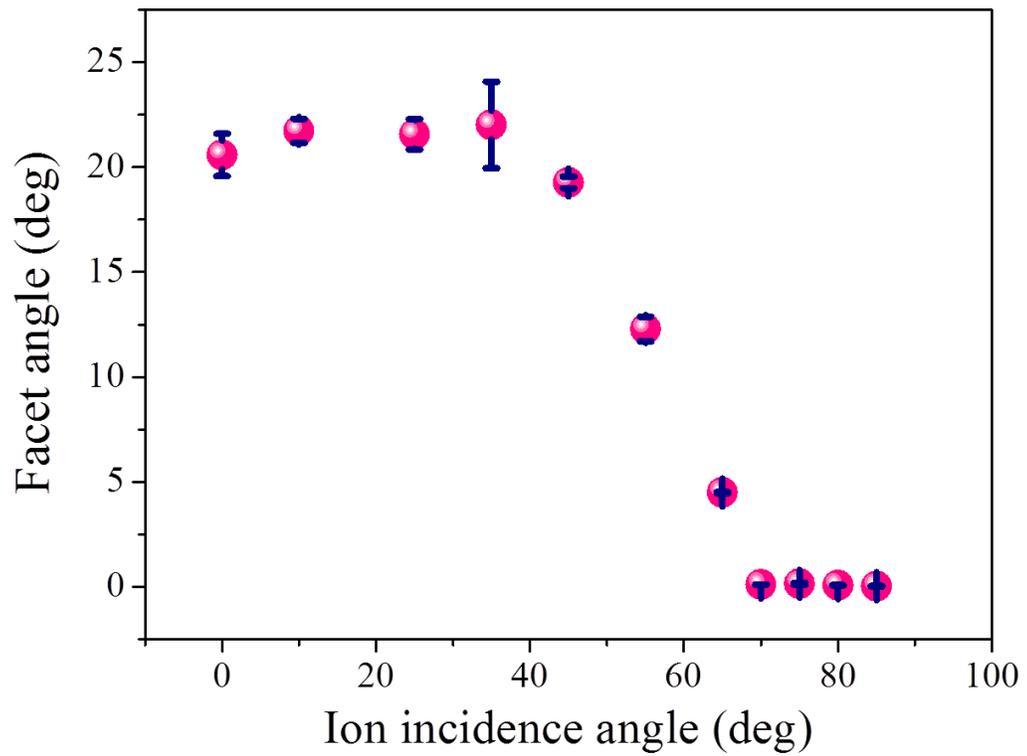

Fig. 7: Evolution of the facet angle extracted from slope angle distributions [cf. Fig. 6], as a function of ion fluence.

1717